\pdfoutput=1
\documentclass[english,11pt,reqno]{article}
\usepackage{lmodern}

\usepackage[T1]{fontenc}
\usepackage{babel}
\usepackage{array}
\usepackage{float}
\usepackage{multirow}
\usepackage{amsmath}
\usepackage{amsthm}
\usepackage{amssymb}
\usepackage{graphicx}
\usepackage[unicode=true,pdfusetitle,
 bookmarks=true,bookmarksnumbered=false,bookmarksopen=false,
 breaklinks=false,pdfborder={0 0 1},backref=false,colorlinks=false]
 {hyperref}

\makeatletter

\providecommand{\tabularnewline}{\\}

\numberwithin{equation}{section}
\numberwithin{figure}{section}
\numberwithin{table}{section}

\usepackage{graphicx}
\usepackage{fancyhdr}
\usepackage{geometry}

\usepackage{fancybox}
\usepackage{lastpage}
\usepackage{longtable}

\usepackage{longtable}
\usepackage{subfigure}
\newcommand{\bee}{\begin{eqnarray}} 
\newcommand{\eee}{\end{eqnarray}}
\newcommand{\be}{\begin{equation}} 
\newcommand{\ee}{\end{equation}}

\usepackage{makecell}
\usepackage{placeins}
\numberwithin{equation}{section}

\numberwithin{figure}{section}

\numberwithin{table}{section}

\AtBeginDocument{
  
}

\makeatother

\begin{document}

\title{Deep Learning-Based Least Square Forward-Backward Stochastic
Differential Equation Solver for High-Dimensional Derivative Pricing}

\author{Jian Liang\thanks{Corporate Model Risk, Wells Fargo Bank, jian.liang@wellsfargo.com},
Zhe Xu\thanks{Corporate Model Risk, Wells Fargo Bank, zhe.xu@wellsfargo.com},
Peter Li\thanks{Corporate Model Risk, Wells Fargo Bank, peter.li@wellsfargo.com}}

\date{v1.1 released June 2020}
\maketitle
\begin{abstract}
We propose a new forward-backward stochastic differential equation
solver for high-dimensional derivative pricing problems by combining
deep learning solver with least square regression technique widely
used in the least square Monte Carlo method for the valuation of American
options. Our numerical experiments demonstrate the accuracy of our least square 
backward deep neural network solver and
its capability to produce accurate prices for complex early exercise
derivatives, such as callable yield notes. Our method can serve as
a generic numerical solver for pricing derivatives across various
asset groups, in particular, as an accurate means for pricing high-dimensional
derivatives with early exercise features.
\end{abstract}


\tableofcontents{}

\section{Introduction}

As explained in the well-known book by \cite{JohnHull2018}, a financial derivative, or simply a derivative, can be  defined as a financial instrument whose value depends on (or derives from) the values of  other, more basic underlying variables. The underlying variables can be traded assets such as common stocks, commodities, and bonds, etc. They can also be stock indexes, interest rates, foreign exchange rates, etc.  Derivatives can be used by hedgers to reduce the risk that they face from potential future movements in a market variable and by speculators to bet on the future direction of a market variable.
The most common derivative types are futures contracts,
forward contracts, swaps and options. 

Derivative pricing has been widely studied in academia and industry. 
Numerical methods have to be used for derivative pricing 
except for simple derivatives, such as futures, forwards, swaps, and
European vanilla options. Tree, PDE and Monte Carlo are the three major methods
in pricing complex derivatives. However, both the tree  approach and the classical
finite difference based PDE approach are infeasible for high-dimensional derivative pricing due to the implementation complexity
and the numerical burden. This is the well-known `curse of dimensionality'.
Therefore, Monte Carlo method is widely used in high-dimensional derivative
pricing. In order to determine the optimal exercise strategy, some additional numerical procedures have to be embedded in
Monte Carlo method when pricing early exercisable products, e.g. American
options, Bermudan options, callable structured notes, etc. Since it is computationally impractical to perform a sub-MC simulation at the early exercise time to compute the expected payoff from continuation, various techniques have been  proposed. \cite{Barraquand} proposed a stratified state method which
sorts the stock price paths according to a state variable (rather
than the stock price) to determine the payoff. However, in Barraquand and Martineau's method, an error estimate of the results cannot be obtained. 
\cite{Broadie} proposed a simulated tree method to price American options, and their approach can also generate
the upper and lower bounds for American options. \cite{LSQ_MC} proposed a least square Monte Carlo algorithm to
price American options, and in their approach a least square regression
was introduced in the early exercise step to estimate the expected payoff from continuation. Since it is computationally efficient and easy to be implemented (\cite{Stentoft2004}), the
least square Monte Carlo is the most widely used algorithm among practitioners
for pricing high-dimensional derivatives with early exercise features. 

The application of machine learning in derivative pricing can be traced back to as early as 1990s when \cite{Hutchinson1994} used neutral network in a nonparametric regression to estimate the option prices. 
In a recent study \cite{De2018} applied machine leaning in Gaussian Process Regression to predict option prices from the neutral network trained by product, market and model parameters.  Alternative to using them as regression tools in the context of option pricing, more recently, researchers have used machine learning techniques to approximate the solutions to parabolic PDEs associated with derivative pricing, in particular, for high-dimensional PDEs where classical approaches find challenges.  
\cite{SirignanoSiam2017, SirignanoJcp2018} 
applied the deep Galerkin method to solve high-dimensional PDEs that arise in quantitative
finance applications including option pricing. \cite{E2017,Han8505}
proposed an innovative algorithm, referred as forward DNN in this paper, where the deep neural network is used to solve non-linear parabolic PDE. Through the generalized Feynman-Kac theorems they formulated the PDE into
equivalent backward stochastic differential equations (BSDEs), then
developed a deep neutral network algorithm to solve the BSDEs. Their algorithm is straightforward to implement and can be directly applied in European style high-dimensional
derivative pricing. \cite{Raissi2018} proposed a different
loss function from that in the work of \cite{E2017} and placed the neutral network directly  on the solution of the interests. As a result, in Raissi's algorithm,  the solution covers the whole space-time domain, not just the initial point as what is in Weinan E's algorithm. In addition, Raissi's algorithm enjoys the merit of the independence between the number of parameters of the neural network and the number of time discretization spacing. 
Note that Weinan E's or Raissi's method is more appropriate in European
style derivative pricing, but not for derivatives with early exercise
features. \cite{Fujii2019} demonstrated that the use
of asymptotic expansion as prior knowledge in the forward DNN method
could drastically reduce the loss function and accelerate the convergence
speed. They also extended the forward DNN method for reflected BSDEs
which could be used in American basket option pricing. \cite{Haojie} 
proposed a backward DNN algorithm for pricing Bermudan
swaptions under LIBOR market model. However, the validity and accuracy of their
backward DNN for Bermudan swaptions is not clear since there were no numerical
studies in the work of Wang {\itshape et al.} to compare the results from the
backward DNN with those from the classical approaches such as the least
square Monte Carlo simulation. Since the discounted payoff on a simulation path is taken as the continuation value conditional on no early exercise, the price of a Bermudan swaption is expected to be biased in Wang {\itshape et al.}'s algorithm. 

In this paper, we propose a backward deep learning-based least square forward-backward
stochastic differential equation solver for pricing high-dimensional
derivatives, in particular, with early exercise features. The application
of neutral networks combined with regression to tackle early exercise
options, such as American options pricing problems, has been reported
by \cite{doi:10.1111/j.1467-9965.2010.00404.x}. In Kohler
{\itshape et al.}'s work neutral network was used as an optimization tool for
non-parametric regression, while in our work neutral network is used
to solve the BSDE. Different from Wang {\itshape et al.}'s algorithm, which is only applicable to a 
vanishing drift term, our algorithm can be used for general drift functions. 
In addition, in our algorithm the least
square regression is used to determine the optimal condition for early
exercises. Even though there have been many researches on using neural
network to approximate the solutions of PDEs for the purpose of derivative
pricing, very little studies have been reported to assess its efficiency
compared to classical numerical methods. Our work also aims
at closing this gap by comparing the DNN based algorithms with the classical
Monte Carlo simulation and hence providing guidance on what situations
DNN based algorithms are more efficient. 

The rest of this paper is organized as follows. In Section \ref{sec:Background-Knowledge},
we introduce some basic background knowledge for forward-backward
stochastic differential equations (FBSDEs), which is the key knowledge to
our least square backward DNN method. In addition, we briefly explain the concept of
Bermudan options and callable yield notes, which will be used as examples
in our numerical testing. The forward DNN method (\cite{E2017}) is
described in Section \ref{sec:Forward-DNN-Method}. In Section \ref{sec:Least-Square-Backward},
we first outline the backward DNN method, and then introduce the least
square backward DNN method. Numerical results for Bermudan options
and callable yield notes are presented in Section \ref{sec:Numerical-Results}. Section \ref{sec:Numerical-Results} also includes the efficiency tests of DNN based methods.
We conclude our paper in Section \ref{sec:Conclusion}.

\section{Background knowledge\label{sec:Background-Knowledge}}

In this section, we first introduce some basics of forward-backward
stochastic differential equations (FBSDEs) and then describe the contract characteristics of
Bermudan options and callable yield notes (CYN). Both instruments are used
to perform our numerical tests in Section \ref{sec:Numerical-Results}.

\subsection{Forward-backward stochastic differential equation}

Many pricing and optimization problems in financial mathematics can
be reformulated in terms of backward stochastic differential equations
(BSDEs). These equations are non-anticipatory terminal value problems
for stochastic differential equations (SDEs) of the form
\begin{eqnarray}
-dY_{t} & = & f\left(t,Y_{t},Z_{t}\right)dt-Z_{t}dW_{t},\nonumber \\
Y_{T} & = & \xi,\label{eq:BSDE}
\end{eqnarray}
where $W_{t}$ is a standard $d$-dimensional Brownian motion defined
on a complete probability space. The square-integrable terminal condition
$\xi$ (measurable with respect to filtration generated up to time
$T$ by the Brownian motion) and the so-called drift term $f$ are
given. 

When BSDEs are used in derivative pricing, $Y_{t}$ corresponds
to the derivative value and $Z_{t}$ is related to the hedging portfolio.
In many portfolio optimization problems, $Y_{t}$ corresponds to the
value process, while an optimal control can often be derived from $Z_{t}$.
Finally, BSDEs can also be applied in order to obtain Feynman-Kac
type representation formulas for nonlinear parabolic PDEs. In the equation above, $Y_{t}$
and $Z_{t}$ correspond to the solution and the gradient of the PDE,
respectively. 

In this paper, we will focus on a forward-backward stochastic differential
equation (FBSDE) of the form
\begin{eqnarray}
dX_{t} & = & \mu\left(t,X_{t}\right)dt+\sigma\left(t,X_{t}\right)dW_{t},\nonumber \\
X_{0} & = & x,\nonumber \\
-dY_{t} & = & f\left(t,X_{t},Y_{t},Z_{t}\right)dt-Z_{t}dW_{t},\nonumber \\
Y_{T} & = & g\left(X_{T}\right).\label{eq:FBSDE}
\end{eqnarray}
Here $g(\cdot)$ is the payoff function. The name forward-backward comes from the fact that $X$ moves forward as its
initial value is given, $Y$ moves backward as its terminal value
is given. Suppose $X_{t}$ is the stock value and it follows
\begin{equation}
dX_{t}=\left(r-q\right)X_{t}dt+\sigma X_{t}dW_{t}.\label{eq:stock_SDE}
\end{equation}
For simplicity, we assume $r$ is the constant discount rate, $q$
is the constant dividend,  and $\sigma$ is the constant volatility.
We only use subscript when it is necessary. Proceeding in the same
fashion as in the derivation of the Black-Scholes PDE, we construct
a portfolio $\Pi=Y-\Delta X$, and $\Delta$ will be selected so that
the value of the portfolio is deterministic.
\[
d\Pi=dY-\Delta dX-q\Delta Xdt=\left(\frac{\partial Y}{\partial t}+\frac{1}{2}\sigma^{2}X^{2}\frac{\partial^{2}Y}{\partial X^{2}}\right)dt+\frac{\partial Y}{\partial X}dX-\Delta dX-q\Delta Xdt
\]
The term $q\Delta Xdt$ arises since the stock pays dividend which
decreases the value of the portfolio by the amount of the dividend.
If we select $\Delta=\frac{\partial Y}{\partial X}$, we then have
\[
d\Pi=\left(\frac{\partial Y}{\partial t}+\frac{1}{2}\sigma^{2}X^{2}\frac{\partial^{2}Y}{\partial X^{2}}\right)dt-q\Delta Xdt.
\]
Since the value of the portfolio is risk free, we must have
\[
d\Pi=r\Pi dt=r\left(Y-\Delta X\right)dt.
\]
This leads to the following Black Scholes PDE
\begin{equation}
\frac{\partial Y}{\partial t}+\frac{1}{2}\sigma^{2}X^{2}\frac{\partial^{2}Y}{\partial X^{2}}+\left(r-q\right)X\frac{\partial Y}{\partial X}-rY=0.\label{eq:BS_PDE}
\end{equation}
From It\^{o}'s Lemma, we have
\begin{eqnarray*}
dY & = & \left(\frac{\partial Y}{\partial t}+\frac{1}{2}\sigma^{2}X^{2}\frac{\partial^{2}Y}{\partial X^{2}}\right)dt+\frac{\partial Y}{\partial X}dX\\
 & = & \left(rY-\left(r-q\right)X\frac{\partial Y}{\partial X}\right)dt+\frac{\partial Y}{\partial X}\left(\left(r-q\right)Xdt+\sigma XdW\right)\\
 & = & rYdt+\sigma X\frac{\partial Y}{\partial X}dW,
\end{eqnarray*}
or
\begin{equation}
-dY=-rYdt-\sigma X\frac{\partial Y}{\partial X}dW,\label{eq:dY}
\end{equation}
that is $f=-rY$, and $Z=\sigma X\frac{\partial Y}{\partial X}$ in
Eq \eqref{eq:FBSDE}.

The above statement can be easily extended to a high-dimensional derivative
pricing ($Y=Y\left(X^{1},X^{2},\cdots,X^{d}\right)$), and we have
(neglecting subscript $t$)
\begin{eqnarray}
dX^{i} & = & \mu^{i}\left(t,X^{i}\right)dt+\sigma^{i}\left(t,X^{i}\right)dW^{i}\nonumber \\
X_{0}^{i} & = & x^{i}\nonumber \\
-dY & = & -rYdt-\sum\sigma^{i}X^{i}\frac{\partial Y}{\partial X^{i}}dW^{i}\label{eq:multi_FBSDE}\\
Y_{T} & = & g\left(X_{T}^{1},X_{T}^{2},\cdots,X_{T}^{d}\right)\nonumber \\
cov\left(dW^{i},dW^{j}\right) & = & \rho^{ij}dt\mbox{ }\mbox{ }\mbox{ }\mbox{ }|\rho^{ij}|<1.\nonumber 
\end{eqnarray}

\subsection{Bermudan options}

A Bermudan option is a type of exotic option that can only be exercised
on predetermined dates. The Bermudan option is exercisable on the
date of expiration, and on certain specified dates that occur between
the purchase date and the date of expiration. A Bermudan option can be considered as a
hybrid of an American option (exercisable on any dates before and including
expiration) and a European option (exercisable only at expiration).
The payoff function of a Bermudan call at expiration if not exercised
early is given by
\begin{equation}
V\left(T\right)=\max\left(\sum_{i=1}^{d}\omega_{i}X^{i}\left(T\right)-K,0\right),
\end{equation}
where $K$ is the strike of the option and weights $\omega_{i}$ are
given constants. When an exercise event happens, the option expires
and the holder will receive its intrinsic value. Given the exercise
times as $t<t_{1}<t_{2}<....<t_{n}\leq T$, the value of a Bermudan
option at time $t$ can be written as

\begin{equation}
V(t)=D(t,T)\underset{\tau\in\mathrm{T(t)}}{\sup}E^{Q}\left[V\left(\tau\right)|\mathcal{F}_{t}\right],
\end{equation}
where $D(t,T)$ is the discount factor, $\mbox{T}(t)$ is the set
of exercise times, and the expectation is taken under risk neutral
measure.

\subsection{Callable yield notes}

Callable yield note (CYN), also called worst of issuer callable, is
a yield enhancement product. The performance of a CYN is capped by a coupon
that is guaranteed by an issuer. As the name implies, the issuer,
at his discretion, can call the product, usually on predefined observation
dates. The underlying entities are generally composed of several stocks
or stock indices; thus making it a product based on a worst-of function.
The call notice dates for a CYN are often identical to the coupon
record dates. We denote the coupon record dates as $t_{i}$, $i=1,2,\cdots,N$,
with $t_{N}=T$ being equal to the expiry date $T$. The coupon payments
are subject to a barrier condition and the knock-in barrier is observed
at expiry. The coupon payments per unit of notional are
\begin{eqnarray}
c\left(t_{i}\right) & = & r_{i}\Theta\left(p\left(t_{i}\right)-B_{i}\right)\mbox{ }\mbox{ }\mbox{ }\mbox{ }\mbox{ }\mbox{ for }i=1,2,\cdots,N-1\nonumber \\
c\left(t_{N}\right) & = & r_{N}\Theta\left(p\left(T\right)-B_{N}\right)-\Theta\left(B-p\left(T\right)\right)\max\left(K-p\left(T\right),0\right),
\end{eqnarray}
where $r_{i}$ is the contingent coupon with coupon barrier $B_{i}$
on $i$th coupon day, $B$ is the knock-in barrier at expiry, $K$
is the knock-in put strike, and $p\left(t\right)$ is the relevant
performance since trade inception. $p\left(t\right)$ is defined as
\begin{equation}
p\left(t\right)=\min_{j\in\left\{ 1,2,\cdots,d\right\} }\left[\frac{X^{j}\left(t\right)}{X^{j}\left(0\right)}\right],\label{eq:performance}
\end{equation}
and $\Theta\left(x\right)$ is the Heaviside function
\begin{equation}
\Theta\left(x\right)=\begin{cases}
1 & \mbox{ }\mbox{ for }x\geq0\\
0 & \mbox{ }\mbox{ otherwise}.
\end{cases}\label{eq:heaviside}
\end{equation}
Furthermore, upon redemption (at the scheduled expiry or early issuer
call) the principal notional is returned to the holder. That is
\begin{eqnarray}
payoff\left(T\right) & = & notional+c\left(t_{N}\right)\nonumber \\
callvalue\left(t_{i}\right) & = & notional\mbox{ }\mbox{ }\mbox{ }\mbox{ }\mbox{ for }i=1,2,\cdots,N-1.\label{eq:call}
\end{eqnarray}
Given the call times as $t<t_{1}<t_{2}<....<t_{n}\leq T$, the value
of a callable yield note at time $t$ is
\begin{equation}
V(t)=D(t,T)\underset{\tau\in\mathrm{T(t)}}{\inf}E^{Q}\left[V\left(\tau\right)|\mathcal{F}_{t}\right],
\end{equation}
where $D(t,T)$ is the discount factor, $\mbox{T}(t)$ is the set
of call times, and the expectation is taken under risk neutral measure.

\section{Forward DNN method\label{sec:Forward-DNN-Method}}

Forward solvers using deep neural network (DNN) have been developed
mainly by \cite{E2017} and \cite{Han8505}. FBSDEs (Eq \eqref{eq:FBSDE})
can be numerically solved in the following way:
\begin{itemize}
\item Simulate sample paths for the FBSDE using a standard Monte Carlo method.
\item Approximate $Z$ using a deep neural network (DNN), then plug into
the FBSDE to propagate along time. 
\end{itemize}
We briefly describe the forward DNN method in this section. More details
can be found in \cite{E2017} and \cite{Han8505}. For simplicity, we use
1D (single underlier) case as an example. The high-dimensional case is
similar. Specifically, consider a time discretization $\pi=\left\{ t_{0},\cdots,t_{N}\right\} $
of the interval $[0,T]$, i.e. $0=t_{0}<t_{1}<\cdots<t_{N}=T$, where
we assume valuation date=0 and expiration date$=T$. Denoting $h_{i}=t_{i+1}-t_{i}$
and $dW_{i}=W_{t_{i+1}}-W_{t_{i}}$. 
\begin{enumerate}
\item $M$ Monte Carlo (MC) paths of the underlying stock $X_{i}$ (short
for $X_{t_{i}}$, similarly for other notations) are sampled by an
Euler scheme through
\begin{equation}
X_{i+1}=X_{i}+\mu\left(t_{i},X_{i}\right)h_{i}+\sigma\left(t_{i},X_{i}\right)dW_{i}.\label{eq:X_SDE}
\end{equation}
This step is the same as the standard Monte Carlo pricer. Other discretization
schemes can be used, for instance, log-Euler discretization or Milstein
discretization (\cite{Milstein}). 
\item At time $t_{0}=0$, $Y_{0}$ and $Z_{0}$ are randomly picked. 
\item For $t_{i}\in\pi$, we have
\begin{equation}
Y_{i}-Y_{i+1}=f(t_{i},X_{i},Y_{i},Z_{i})h_{i}-Z_{i}dW_{i},
\end{equation}
or
\begin{equation}
Y_{i+1}=Y_{i}-f(t_{i},X_{i},Y_{i},Z_{i})h_{i}+Z_{i}dW_{i}.
\end{equation}
At each time step $t_{i}$, given $Y_{i}$, a deep neural network
(DNN) approximation is used for $Z_{i}$ as $Z_{i}\left(\theta_{i}\right)$
for some hyper-parameter $\theta_{i}$ using sampled data $X_{i}$.
Then, the FBSDE is propagating forward in time direction from $t_{i}$
to $t_{i+1}$ as
\begin{equation}
Y_{i+1}=Y_{i}-f\left(t_{i},X_{i},Y_{i},Z_{i}\left(\theta_{i}\right)\right)h_{i}+Z_{i}\left(\theta_{i}\right)dW_{i}.\label{eq:EWN_FBSDE}
\end{equation}
Along each Monte Carlo path, as propagating forward from time 0 to
$T$, one can estimate $Y_{N}^{\left(j\right)}$ as $Y_{N}^{\left(j\right)}\left(Y_{0},Z_{0},\mathbb{\theta}^{\left(j\right)}\right)$,
where $\mathbf{\theta}^{\left(j\right)}=\left\{ \theta_{0}^{\left(j\right)},\cdots,\theta_{N-1}^{\left(j\right)}\right\} $
are all hyper-parameters for neural network at each time step for
the $j$th MC path.
\item A natural loss function will be
\begin{equation}
L_{Forward}=Mean_{all\ paths}\left(Y_{N}^{\left(j\right)}\left(Y_{0},Z_{0},\mathbf{\theta}^{\left(j\right)}\right)-g\left(X_{N}^{\left(j\right)}\right)\right)^{2}.\label{eq:EWN_Loss}
\end{equation}
Here $g(\cdot)$ is the payoff function
 
\item The Adam optimization (in TensorFlow library) is used to minimize
the loss function $L_{Forward}$ and estimate $Y_{0}$ as
\begin{equation}
\widetilde{Y_{0}}=\arg\min_{Y_{0}\mbox{ \ensuremath{}\ensuremath{}\ensuremath{}}}Mean_{all\ paths}\left(Y_{N}^{\left(j\right)}\left(Y_{0},Z_{0},\mathbf{\theta}^{\left(j\right)}\right)-g\left(X_{N}^{\left(j\right)}\right)\right)^{2}.\label{eq:EWN_solution}
\end{equation}
The estimated $\widetilde{Y_{0}}$ is the desired derivative value
at $t=0$. More details about the use of Adam optimization to solve
the above minimization problem can be found in \cite{E2017} and \cite{Han8505}.
\end{enumerate}

\section{Least square backward DNN method\label{sec:Least-Square-Backward}}

Since the forward DNN method above can not be applied to price options
with early exercise features, such as Bermudan options, \cite{Haojie} 
proposed a backward DNN method to price Bermudan
swaptions under LIBOR market model. As discussed in the introduction of this paper, Wang 
{\itshape et al.}'s algorithm is only applicable to a backward process with a vanishing drift term
($f=0$ in Eq \eqref{eq:FBSDE}) and the Bermudan swaption price can be biased since the discounted payoff along 
a simulation path is taken as the continuation value conditional on not exercised early. 
Different from their algorithm, our backward DNN method can be applied to general
drift functions. We also apply the least square regression to estimate continuation value in order to determine
the optimal exercise decision.

\subsection{Backward DNN method}

We would like to propagate backward in time direction and apply the
call/put and coupon events to the derivative value. From Eq \eqref{eq:EWN_FBSDE},
we have
\begin{equation}
Y_{i}=Y_{i+1}+f\left(t_{i},X_{i},Y_{i},Z_{i}\left(\theta_{i}\right)\right)h_{i}-Z_{i}\left(\theta_{i}\right)dW_{i}.\label{eq:back_propagate}
\end{equation}
As we propagate backward in time direction from $t_{i+1}$ to $t_{i}$,
$Y_{i+1}$ is known while $Y_{i}$ is to be determined. We use 1st
order Taylor expansion to do the approximation.
\begin{equation}
Y_{i}\approx Y_{i+1}+\left(f\left(t_{i},X_{i},Y_{i+1},Z_{i}\left(\theta_{i}\right)\right)-\frac{\partial f}{\partial Y}\left(t_{i},X_{i},Y_{i+1},Z_{i}\left(\theta_{i}\right)\right)\left(Y_{i+1}-Y_{i}\right)\right)h_{i}-Z_{i}\left(\theta_{i}\right)dW_{i},\label{eq:back_1}
\end{equation}
which leads to
\begin{equation}
Y_{i}\approx Y_{i+1}+\frac{1}{1-\frac{\partial f}{\partial Y}\left(t_{i},X_{i},Y_{i+1},Z_{i}\left(\theta_{i}\right)\right)h_{i}}\left(f\left(t_{i},X_{i},Y_{i+1},Z_{i}\left(\theta_{i}\right)\right)h_{i}-Z_{i}\left(\theta_{i}\right)dW_{i}\right).\label{eq:back_2}
\end{equation}
One can use higher order Taylor expansion to achieve more precise
approximation. For our particular equations (Eq \eqref{eq:dY}), 1st
order Taylor expansion approximation is indeed the exact solution.
And we have
\begin{equation}
Y_{i}=\frac{Y_{i+1}-Z_{i}\left(\theta_{i}\right)dW_{i}}{1+rh_{i}}.\label{eq:back_3}
\end{equation}
Starting from $t_{N}=T$, we can propagate backward in time direction
to $t_{0}=0$, and obtain the estimated initial value $Y_{0}^{\left(j\right)}\left(\theta^{\left(j\right)}\right)$
for each sampled path, where $\mathbf{\theta}^{\left(j\right)}=\left\{ \theta_{0}^{\left(j\right)},\cdots,\theta_{N-1}^{\left(j\right)}\right\} $
are all hyper-parameters for neural network at each time steps for
the $j$th MC path. The ideal case will be that the estimated initial
values $Y_{0}^{\left(j\right)}\left(\theta^{\left(j\right)}\right)$
concentrate to one point. Therefore, the loss function is defined
as
\begin{equation}
L_{Backward}=Mean_{all\ paths}\left(Y_{0}^{\left(j\right)}\left(\theta^{\left(j\right)}\right)-Mean_{all\ paths}\left(Y_{0}^{\left(j\right)}\left(\theta^{\left(j\right)}\right)\right)\right)^{2}.\label{eq:back_loss}
\end{equation}
This implies that we are trying to minimize the variance of the estimated
initial values. The Adam optimization is used to minimize the loss
function $L_{Backward}$ and estimate $Y_{0}$ as
\begin{equation}
\widetilde{Y_{0}}=Mean_{all\ paths}\left(Y_{0}^{\left(j\right)}\left(\widetilde{\theta}^{\left(j\right)}\right)\right),\label{eq:back_solution}
\end{equation}
where
\begin{equation}
\widetilde{\theta}^{\left(j\right)}=\arg\min_{\theta\mbox{ \ensuremath{}\ensuremath{}\ensuremath{}}}L_{Backward}.\label{eq:back_argmin}
\end{equation}
Finally the estimated $\widetilde{Y_{0}}$ is our desired derivative value
at $t=0$.

\subsection{Least square regression}

We use a Bermudan call option to explain how the conditional expectation
of the payoff estimated from least square regression is used to determine
optimal strategy at an early exercise time. The readers are referred
to the classical paper by \cite{LSQ_MC} for more
details. Without loss of generality, we assume the exercise time as $t_{k}\in\pi=\left\{ t_{0},\cdots,t_{N}\right\} $.
The main idea is to employ a regression equation, e.g.,
\begin{equation}
Y_{k}=a+bX_{k}+cX_{k}^{2}+v,\label{eq:LSQ}
\end{equation}
where $v$ is the white noise and $v\sim N\left(0,\eta^{2}\right)$.
The expected derivative value is estimated as
\begin{equation}
\mathbb{E}Y_{k}=a+bX_{k}+cX_{k}^{2}.\label{eq:expectation}
\end{equation}
At an exercise time, the above least square regression is performed
over all the in-the-money paths that have positive call values. Note
that other basis functions can be used in the least square regression,
e.g. weighted Laguerre polynomials, which are used in the paper by
\cite{LSQ_MC}. The optimal strategy at an exercise time can be determined by comparing
the call value (i.e. the immediate exercise value) with the expectation
of the derivative value from continuation
\begin{equation}
Y_{k}=\begin{cases}
Y_{k} & \mbox{if \ensuremath{\mathbb{E}Y_{k}\geq callvalue\left(t_{k},X_{k}\right)}}\\
callvalue\left(t_{k},X_{k}\right) & \mbox{if \ensuremath{\mathbb{E}Y_{k}<callvalue\left(t_{k},X_{k}\right)}}
\end{cases}.\label{eq:call_new}
\end{equation}

\subsection{Least square backward DNN method}
We summarize our least square backward DNN method as follow:
\begin{enumerate}
\item $M$ Monte Carlo (MC) paths of the underlying stock $X_{i}$ (short
for $X_{t_{i}}$, similarly for other notations) are sampled by an
Euler scheme through Eq \eqref{eq:X_SDE}. This step is the same as
forward DNN method. 
\item As we have the sampled $X_{N}^{\left(j\right)}$($j=1,2,\cdots,M$)
available, we can calculate the payoff at expiry for the $j$th
sampled path
\begin{equation}
Y_{N}^{\left(j\right)}=g\left(X_{N}^{\left(j\right)}\right).\label{eq:}
\end{equation}

\item At each time step $t_{i}$, given $Y_{i+1}$, a deep neural network
(DNN) approximation is used for $Z_{i}$ as $Z_{i}\left(\theta_{i}\right)$
for some hyper-parameter $\theta_{i}$ using sampled data $X_{i}$.
Then, the FBSDE is propagating backward using Eq \eqref{eq:back_3}
in time direction from $t_{i+1}$ to $t_{i}$. Along each Monte Carlo
path, as propagating backward from time $T$ to 0, one can estimate
$Y_{0}^{\left(j\right)}$ as $Y_{0}^{\left(j\right)}\left(X_{0},Z_{0},\mathbb{\theta}^{\left(j\right)}\right)$, 
where $\mathbf{\theta}^{\left(j\right)}=\left\{ \theta_{0}^{\left(j\right)},\cdots,\theta_{N-1}^{\left(j\right)}\right\} $
are all hyper-parameters for neural network at each time steps for
the $j$th MC path.
\item During propagating from time $T$ to 0, at exercise time $t_{k}$, the
least square regression is performed and the derivative value at each
path is reset using Eq \eqref{eq:call_new}.
\item Set $L_{Backward}$ as the loss function.
\item The Adam optimization is used to minimize the loss function $L_{Backward}$
and estimate $Y_{0}$ as Eq \eqref{eq:back_solution}. The estimated
$\widetilde{Y_{0}}$ is our desired derivative value at $t=0$. 
\end{enumerate}
Notice that the above least square backward DNN method can be easily
extended to high-dimensional derivative pricing ($Y=Y\left(X^{1},X^{2},\cdots,X^{d}\right)$). 

It is well known that the least square regression approach to exercise 
boundary estimation is suboptimal and produces lower biased prices. Since the least square regression 
is also used in our backward DNN method, the prices produced are lower-biased, same as those from classical least square Monte Carlo (LSQ MC).  
Though, theoretically, the bias can be reduced with the number of regressors going to infinite, in practice, 
increasing the number of regressors  does not always reduce bias, as observed 
by \cite{Stentoft2004} and \cite{Moreno}. Stentoft recommended 2 or 3 orders polynomials to be used in regression, 
in particular under high dimensions, as a trade-off between precision and computing time, 
and to prevent performance deterioration.  Since the goal of this paper is not to assess 
the least square regression performance in terms of polynomials types and orders, 
we use second order monic polynomials as basis functions in our numerical tests (section \ref{sec:Numerical-Results}) for illustration purposes.  
Our testing results indicate that the monic polynomial can produce satisfactory results for the products as evidenced by 
consistency with the results from a finite difference PDE solver.
When the least square backward DNN method is used to price early exercisable products, 
its accuracy will be assessed based on comparisons with PDE and classical 
LSQ MC for up to 3 dimensions, and with classical LSQ MC for higher than 3 dimensions.

\section{Numerical results\label{sec:Numerical-Results}}

In this section, we first use an European option to compare the performance of 
forward DNN and backward DNN methods, then use Bermudan options and CYNs as examples to test
our least square backward DNN method, and compare with PDE and Monte
Carlo results. Our finite difference PDE solver is only implemented
for 1D, 2D and 3D cases\footnote{We implemented the 3D operator splitting method introduced in \cite{3D_PDE}}. 
For the MC solver, we use $M=1,000,000$ sampling
paths to estimate the means. Note that we have the relative differences
between 1M and 500K less than 0.5\%.

The market data setting used in all our testing examples is described
in Table \ref{tab:Market-data-setting}. All of our tested examples
are with $T=1$ and time step $N=100$. Therefore, we have time step
size $h_{i}=0.01$.

\begin{table}[!th]
\noindent \centering{}\caption{Market data setting\label{tab:Market-data-setting}}
\vspace{-0.3cm}
\begin{center}
\resizebox{.99\textwidth}{!}{% %
\begin{tabular}{|c|c|c|c|c|c|c|c|c|c|c|}
\hline 
interest rate $r$ & \multicolumn{10}{c|}{$r=0.01$}\tabularnewline
\hline 
time step $N$ & \multicolumn{10}{c|}{$N=100$}\tabularnewline
\hline 
\hline 
 & stock 1 & stock 2 & stock 3 & stock 4 & stock 5 & stock 6 & stock 7 & stock 8 & stock 9 & stock 10\tabularnewline
\hline 
spot $X_{0}$ & 100 & 150 & 200 & 175 & 125 & 100 & 150 & 200 & 175 & 125\tabularnewline
\hline 
dividend rate $q$ & 0.03 & 0.02 & 0.05 & 0.00 & 0.04 & 0.03 & 0.02 & 0.05 & 0.00 & 0.04\tabularnewline
\hline 
volatility $\sigma$ & 0.2 & 0.3 & 0.25 & 0.24 & 0.15 & 0.2 & 0.3 & 0.25 & 0.24 & 0.15\tabularnewline
\hline 
correlation $\rho$ & \multicolumn{10}{c|}{0.3 for all $\rho_{ij}$}\tabularnewline
\hline 
\end{tabular}}\end{center}
\end{table}

The deep neural network setting in our tests is as follows: each of
the sub-neural network approximating $Z_{i}\left(\theta_{i}\right)$
consists of 4 layers (1 input layer {[}$d$-dimensional{]}, 2 hidden
layers {[}$d+10$-dimensional{]}, and 1 output layer {[}$d$-dimensional{]},
where $d$ is the number of underlying entities). In the test, we
run 5,000 optimization iterations of training and validate the trained
DNN every 100 iterations. This produces 50 results. We use the mean
of the 10 results with the least loss function value as our derivative
value. The MC sampling size is $M=5,000$.

\subsection{Forward vs. backward DNN method}

\begin{figure}[!th]
\noindent \centering{}\includegraphics[width=0.5\textwidth]{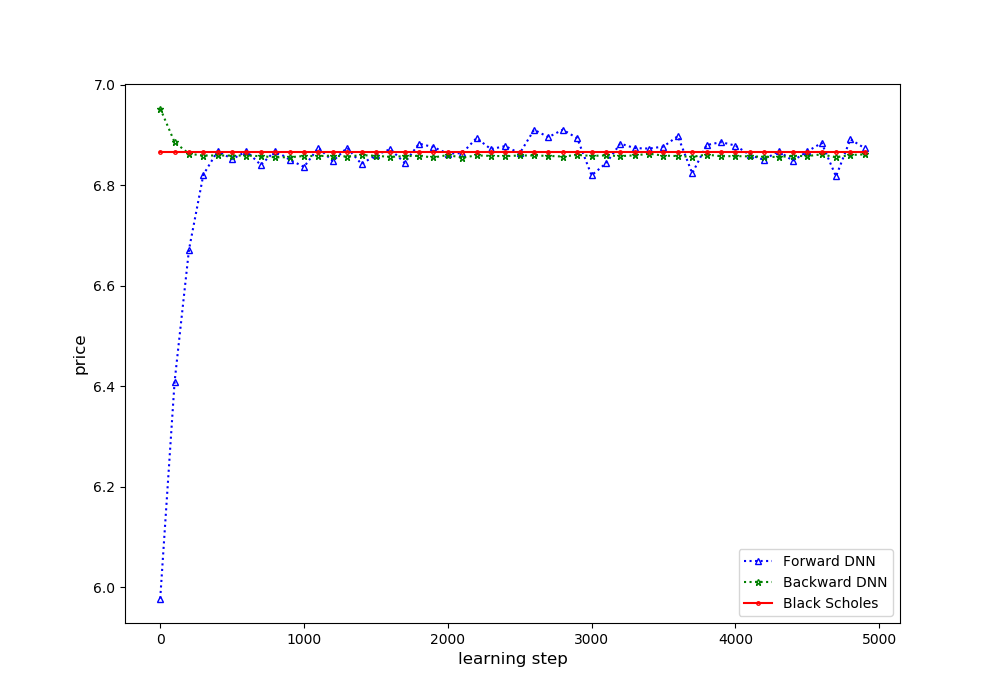}\includegraphics[width=0.5\textwidth]{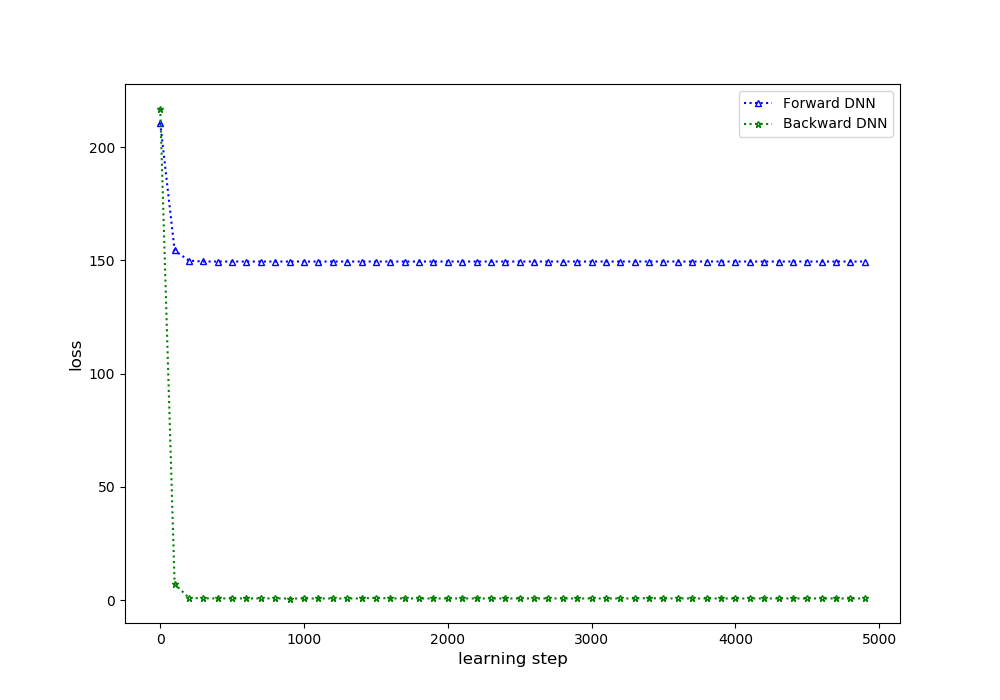}\vspace{-0.5cm}
\caption{Comparison between forward and backward DNN on European call option\label{fig:vanilla}}
\end{figure}

We first use a 1Y ATM single underlying stock European call option
to compare the performance of the forward DNN method with the backward
DNN method. We use stock 1 in Table \ref{tab:Market-data-setting}
as the only underlying stock. The expiration $T=1$ and strike $K=100$.
The results are given in Table \ref{tab:vanilla} and Figure \ref{fig:vanilla}.
Both the forward and the backward DNN methods could provide results close to the Black-Scholes price. Both methods converge fast. Small
oscillations in prices can be observed in the forward DNN method.
By contrast, the Backward DNN method is more stable and converges
slightly faster than the forward DNN method. Therefore, the backward DNN method is the preferred approach.

\begin{table}[!th]
\noindent \centering{}\caption{Comparison between forward and backward DNN on European call option\label{tab:vanilla}}
\begin{tabular}{|c|c|c|c|c|}
\hline 
Black Scholes & \multicolumn{2}{c|}{Forward DNN} & \multicolumn{2}{c|}{Backward DNN}\tabularnewline
\hline 
NPV & NPV & rel diff to BS & NPV & rel diff to BS\tabularnewline
\hline 
\hline 
6.8669  & 6.8688  & 0.03\%  & 6.8575  & -0.14\% \tabularnewline
\hline 
\end{tabular}
\end{table}

\newpage
\subsection{Tests on Bermudan options}\label{sec:Bermudan_Test}

\begin{figure}[!th]
\noindent \centering{}\includegraphics[width=0.5\textwidth]{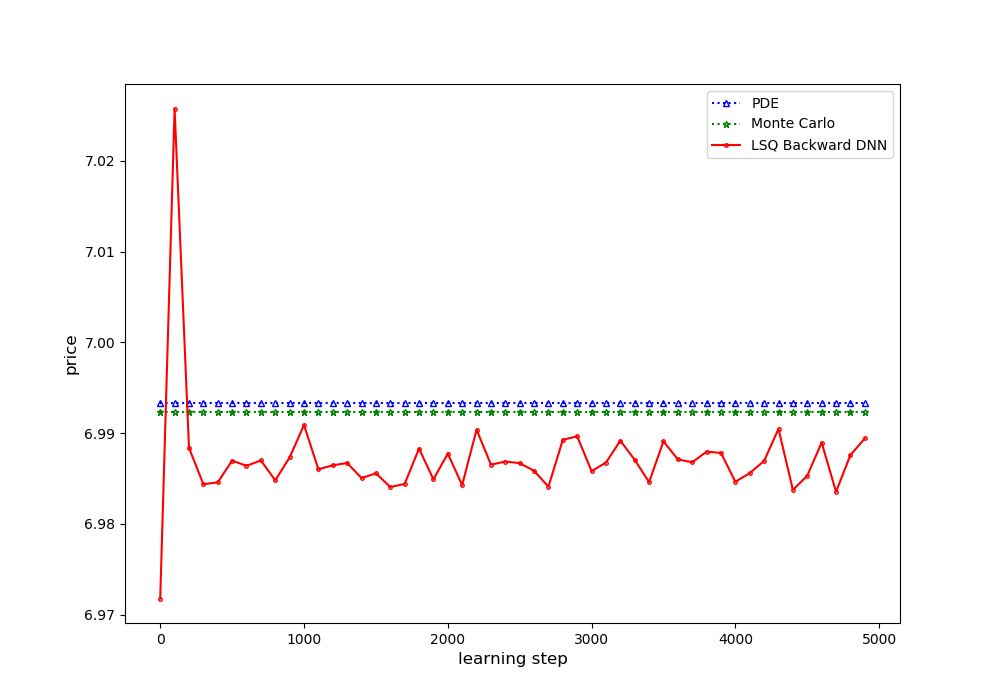}\includegraphics[width=0.5\textwidth]{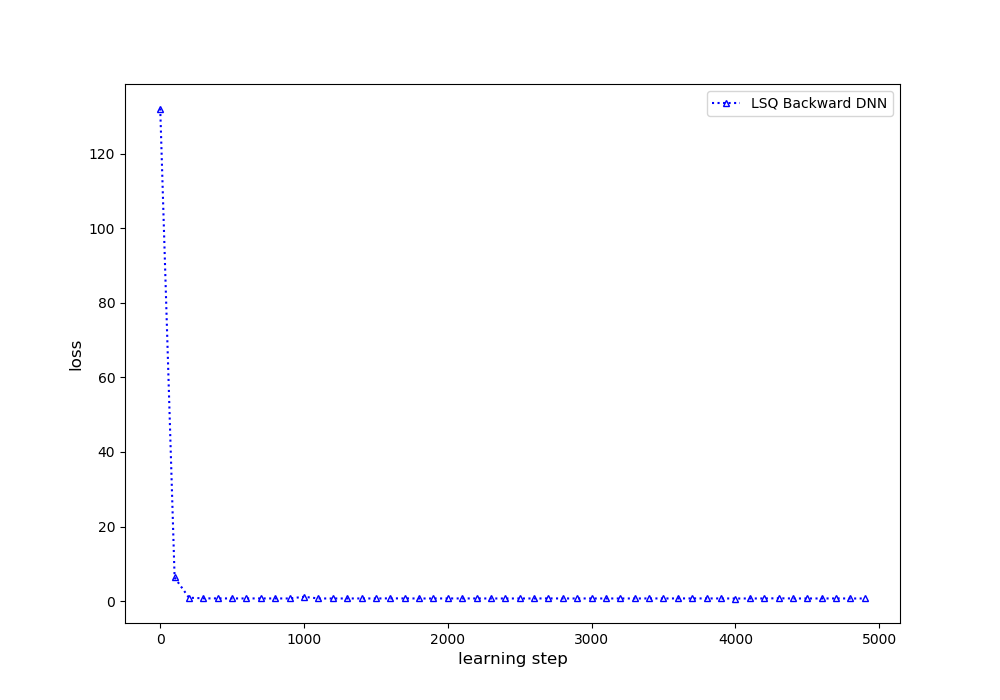}\vspace{-0.5cm}
\caption{Bermudan Call, single underlying stock\label{fig:Bermudan_1UL}}
\end{figure}

\begin{figure}[!th]
\noindent \centering{}\includegraphics[width=0.5\textwidth]{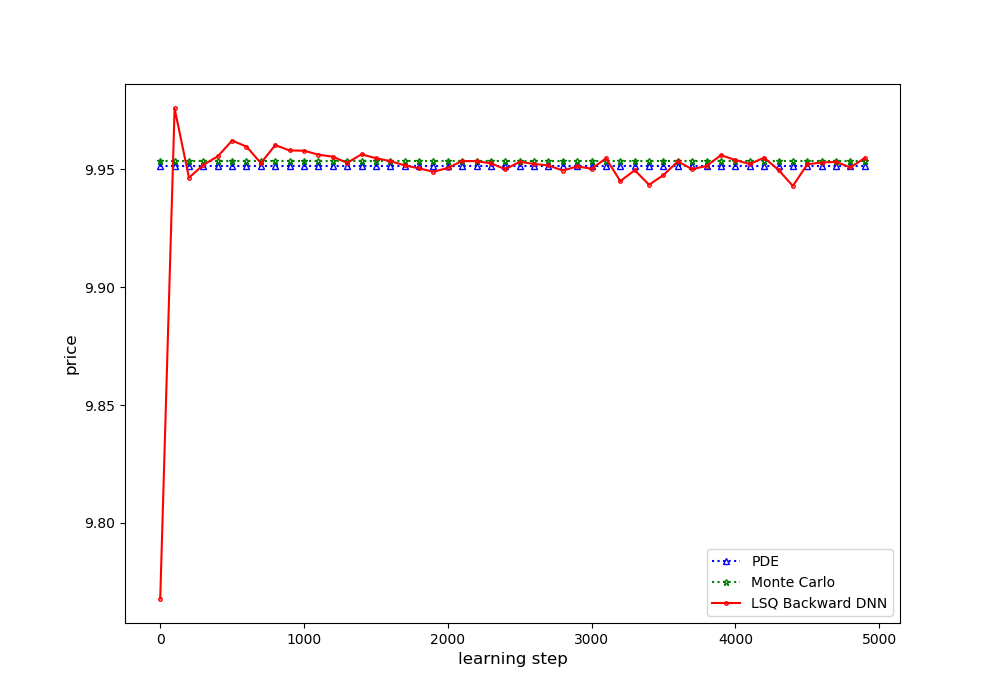}\includegraphics[width=0.5\textwidth]{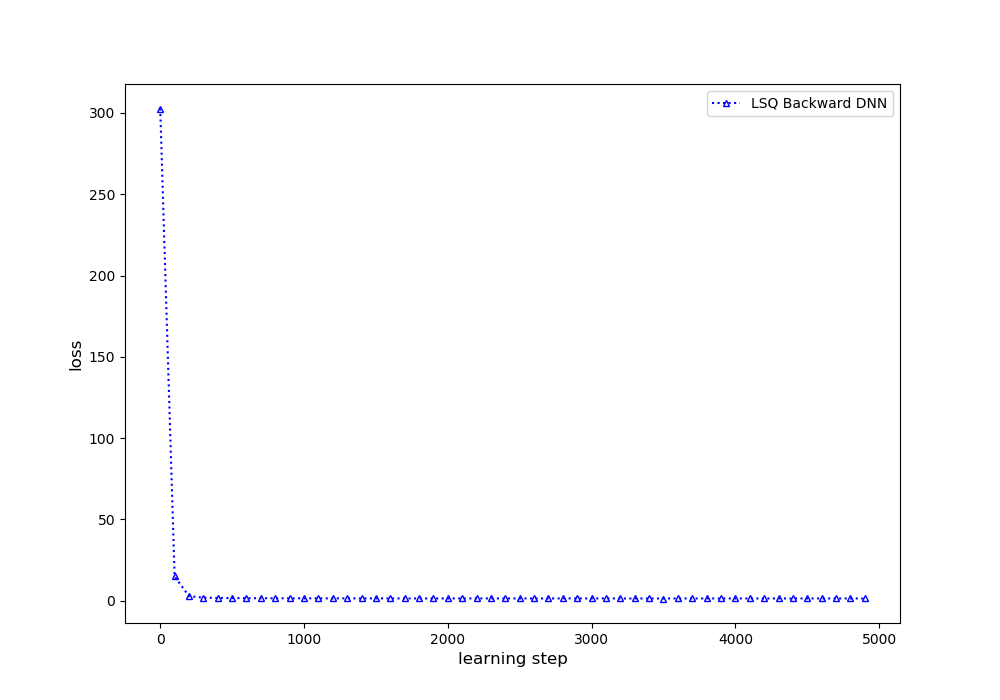}\vspace{-0.5cm}
\caption{Bermudan Call, 2 underlying stocks\label{fig:Bermudan_2UL}}
\end{figure}

\begin{figure}[!th]
\noindent \centering{}\includegraphics[width=0.5\textwidth]{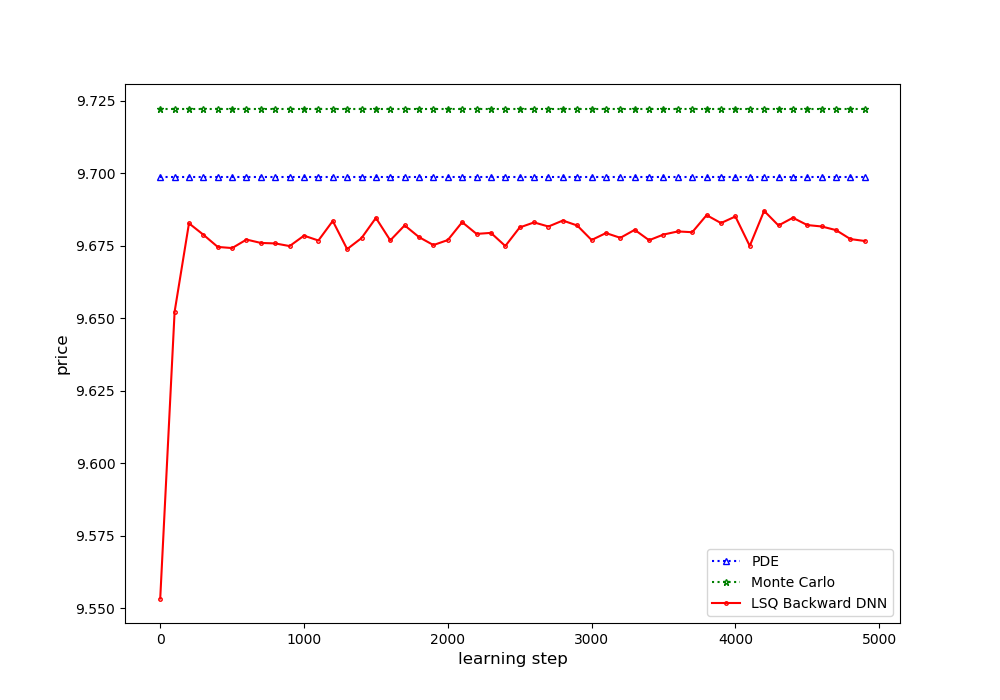}\includegraphics[width=0.5\textwidth]{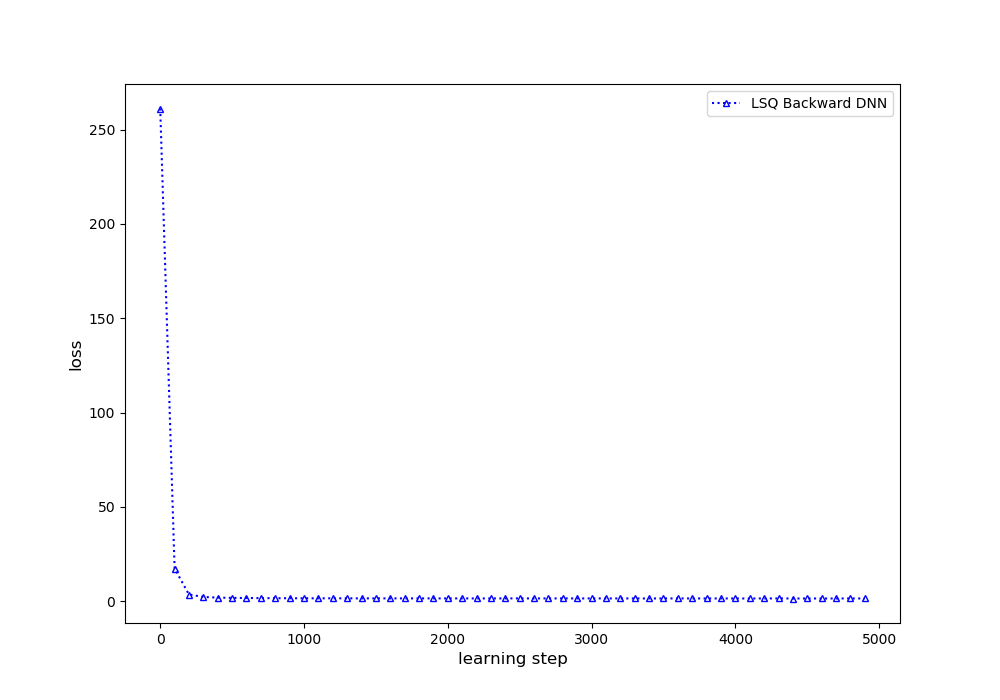}\vspace{-0.5cm}
\caption{Bermudan Call, 3 underlying stocks\label{fig:Bermudan_3UL}}
\end{figure}

\begin{figure}[!th]
\noindent \centering{}\includegraphics[width=0.5\textwidth]{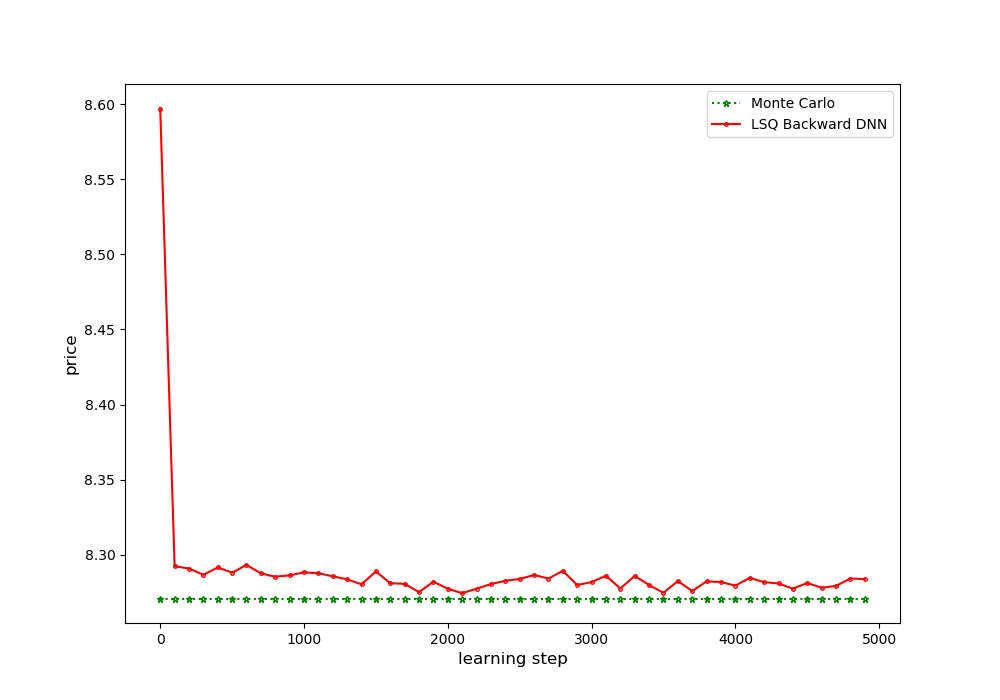}\includegraphics[width=0.5\textwidth]{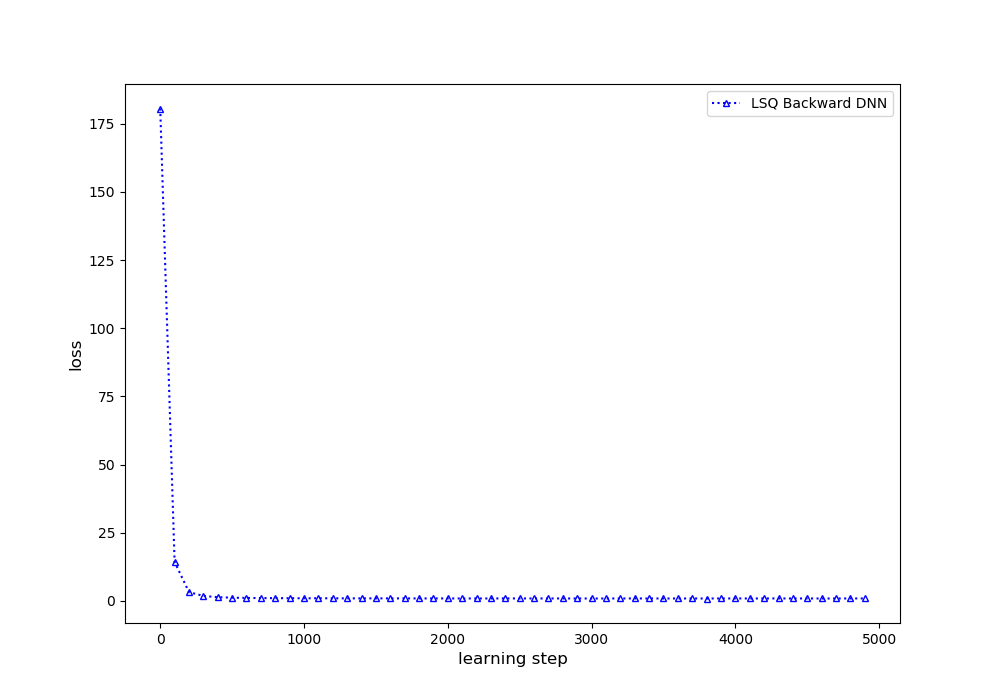}\vspace{-0.5cm}
\caption{Bermudan Call, 5 underlying stocks\label{fig:Bermudan_5UL}}
\end{figure}

In this section, we use 1Y ATM Bermudan options to test the performance
of our least square backward DNN method. We test Bermudan call options
on a single underlying stock, 2 underlying stocks (stock 1, 2), 3
underlying stocks (stock 1, 2, 3) and 5 underlying stocks (stock 1,
2, 3, 4, 5). The strike is chosen as $\sum_{d}\omega_{i}X_{0}^{i}$
with equal weight $\omega_{i}=1/d$ so that the option is ATM. The
Bermuda option can be exercised quarterly, or at $t = 0.25, 0.5, 0.75,
1.0$. We compare the prices from PDE and Monte Carlo with the prices from the least square
backward DNN method. The results are presented in Table \ref{tab:Bermudan}
and Figure \ref{fig:Bermudan_1UL} - \ref{fig:Bermudan_5UL}.
It can be seen that the backward DNN method converges very fast and
the convergence rate is not sensitive to the dimensions of the problem.
The results for 10, 20 and 50 dimensions are presented in Section \ref{sec:eff_test_BermCNY} and 
the largest difference between LSQ MC and the least square backward DNN method 
occurs at 50 dimensions with a difference of 0.4\% (or 2.7 cents).
Overall, the least square backward DNN method can produce very accurate prices for Bermudan call options.

\begin{table}[!th]
\noindent \centering{}\caption{Comparison among PDE, Monte Carlo and least square backward DNN method
on Bermudan options\label{tab:Bermudan}}
\vspace{-0.3cm}
\begin{center}
\resizebox{.99\textwidth}{!}{
\begin{tabular}{|c|c|c|c|c|c|}
\hline 
\multicolumn{1}{|c|}{{\small{}1Y ATM Bermudan Call}} & {\small{}PDE} & {\small{}LSQ MC} & {\small{}LSQ BDNN} & {\small{}rel diff from PDE} & {\small{}rel diff from LSQ MC}\tabularnewline
\hline 
\hline 
{{\small{}single stock (stock 1)}} & {\small{}6.9933} & {\small{}6.9923} & {\small{}6.9863} & {\small{}-0.10\%} & {\small{}-0.09\%}\tabularnewline
\hline 
{{\small{}2 stocks (stock 1, 2)}} & {\small{}9.9514} & {\small{}9.9535} & {\small{}9.9488} & {\small{}-0.03\%} & {\small{}-0.05\%}\tabularnewline
\hline 
{{\small{}3 stocks (stock 1, 2, 3)}} & {\small{}9.6987} & {\small{}9.7224} & {\small{}9.6813} & {\small{}-0.18\%} & {\small{}-0.42\%}\tabularnewline
\hline 
{{\small{}5 stocks (stock 1, 2, 3, 4, 5)}} & {\small{} } & {\small{}8.2709} & {\small{}8.2795} & {\small{} } & {\small{}0.10\%}\tabularnewline
\hline 
\end{tabular}
}
\end{center}
\end{table}

\subsection{Tests on callable yield notes}\label{sec:CYN_test}

\begin{figure}[!th]
\noindent \centering{}\includegraphics[width=0.5\textwidth]{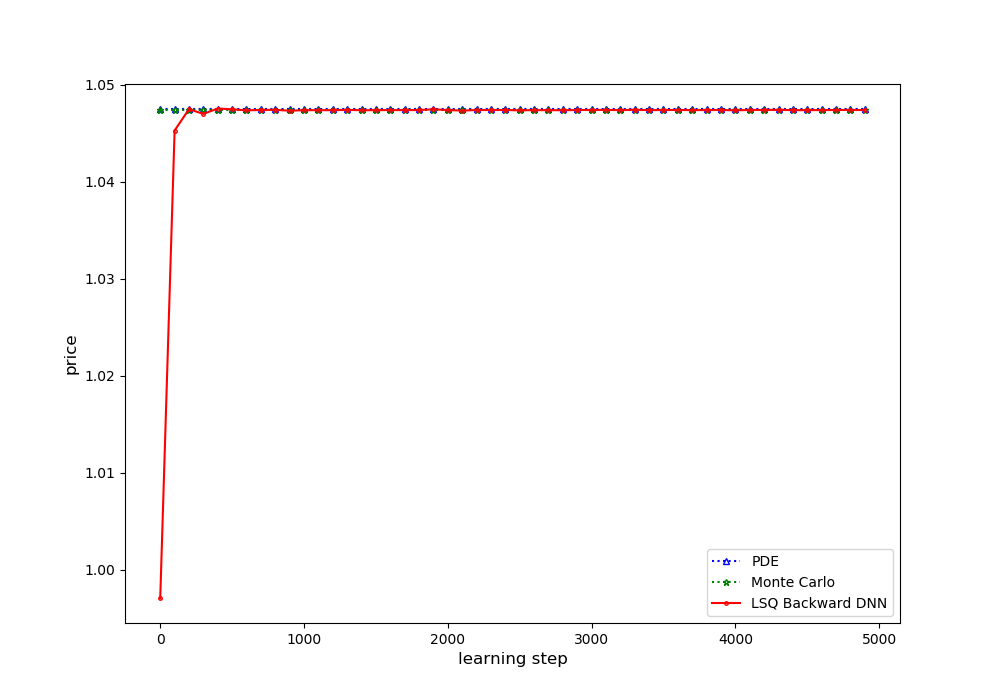}\includegraphics[width=0.5\textwidth]{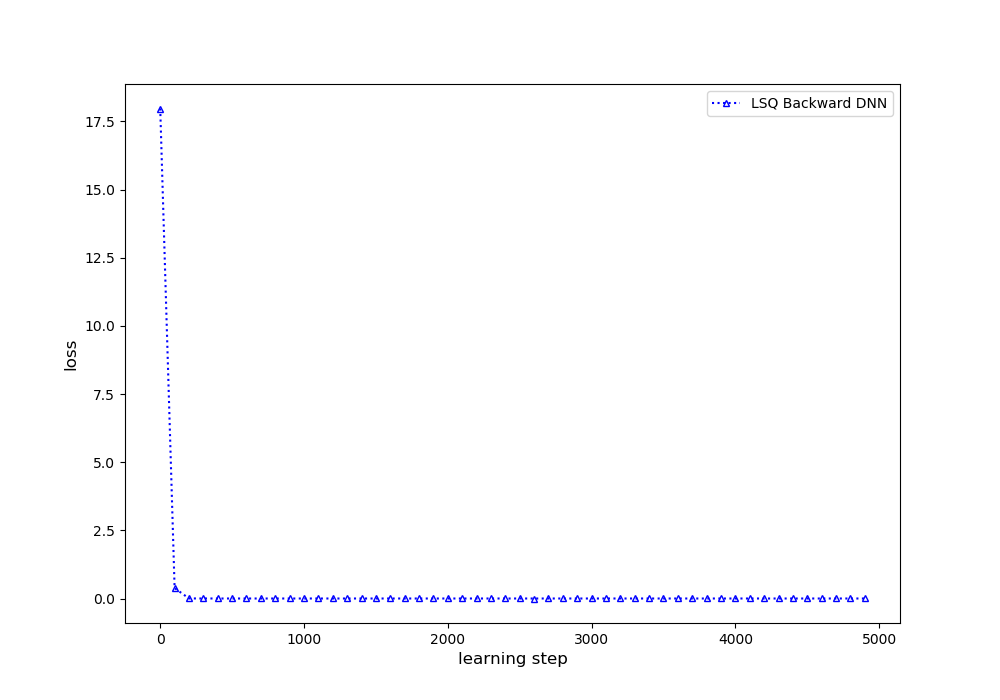}\vspace{-0.5cm}
\caption{CYN, single underlying stock\label{fig:CYN_UL1}}
\end{figure}

\begin{figure}[!th]
\noindent \centering{}\includegraphics[width=0.5\textwidth]{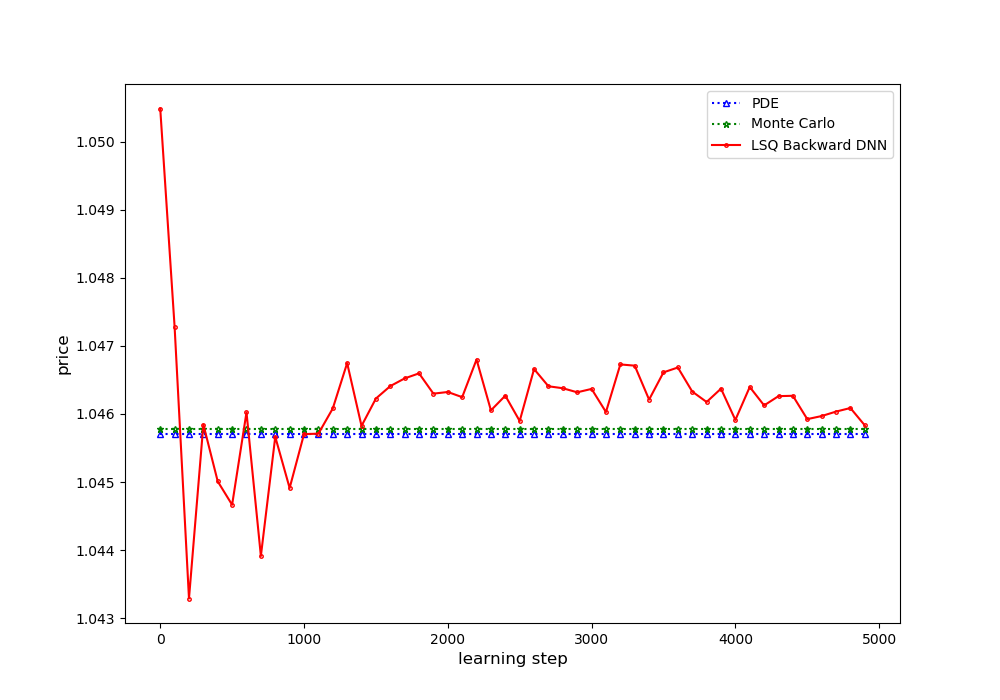}\includegraphics[width=0.5\textwidth]{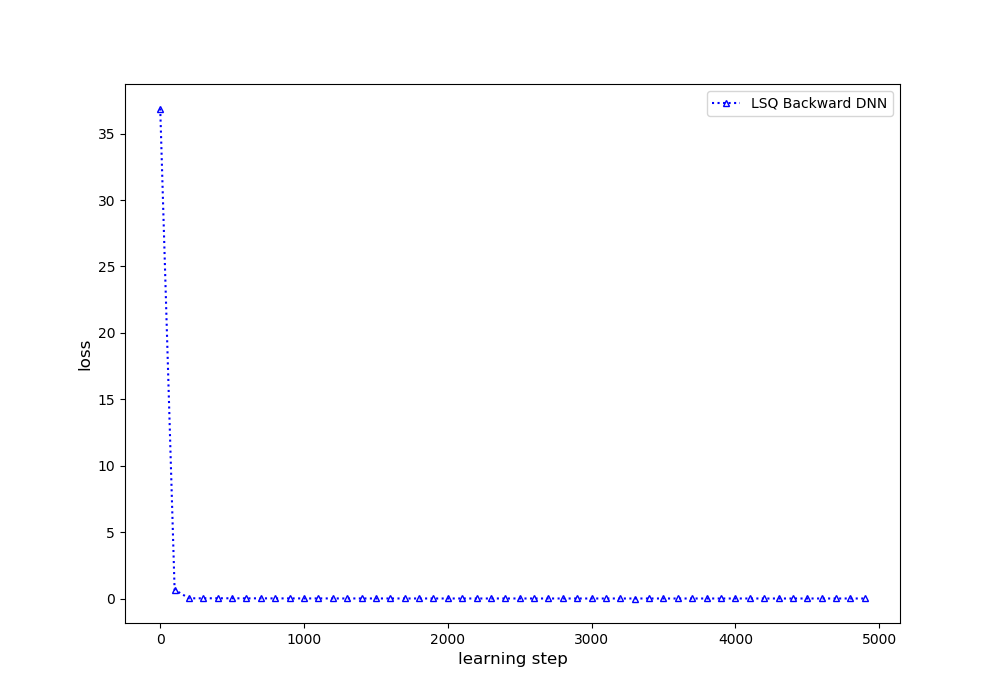}\vspace{-0.5cm}
\caption{CYN, 2 underlying stocks\label{fig:CYN_UL2}}
\end{figure}

\begin{figure}[!th]
\noindent \centering{}\includegraphics[width=0.5\textwidth]{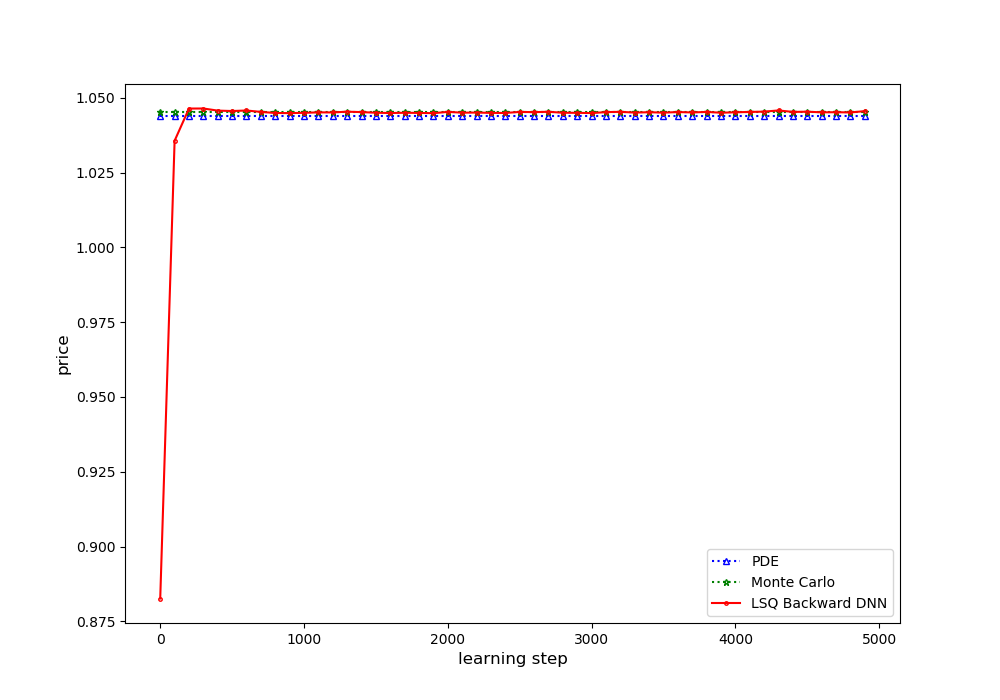}\includegraphics[width=0.5\textwidth]{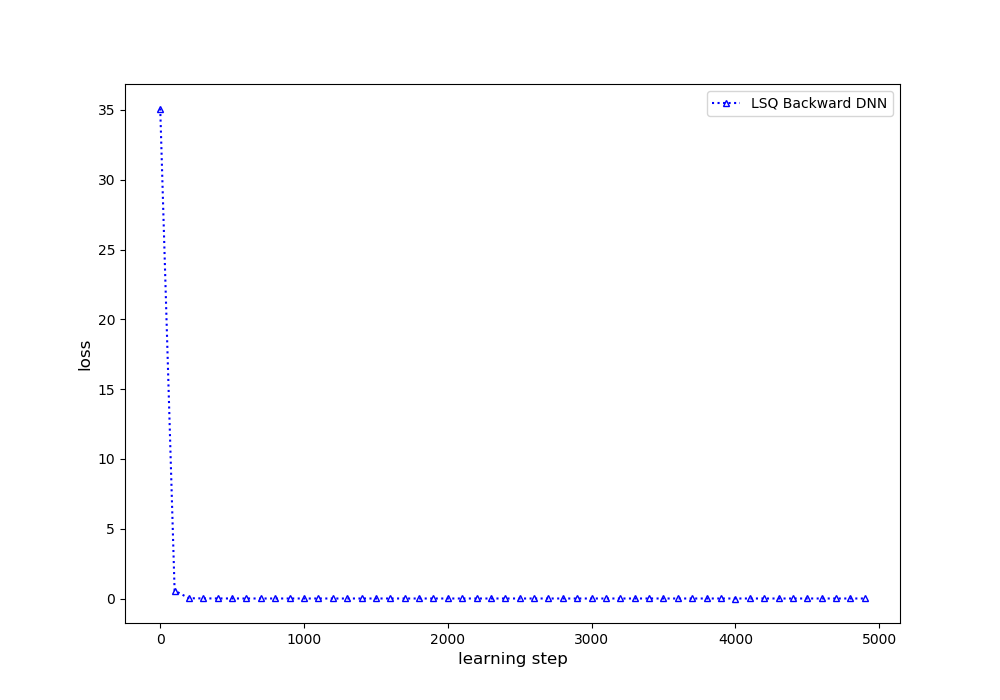}\vspace{-0.5cm}
\caption{CYN, 3 underlying stocks\label{fig:CYN_UL3}}
\end{figure}

\begin{figure}[!th]
\noindent \centering{}\includegraphics[width=0.5\textwidth]{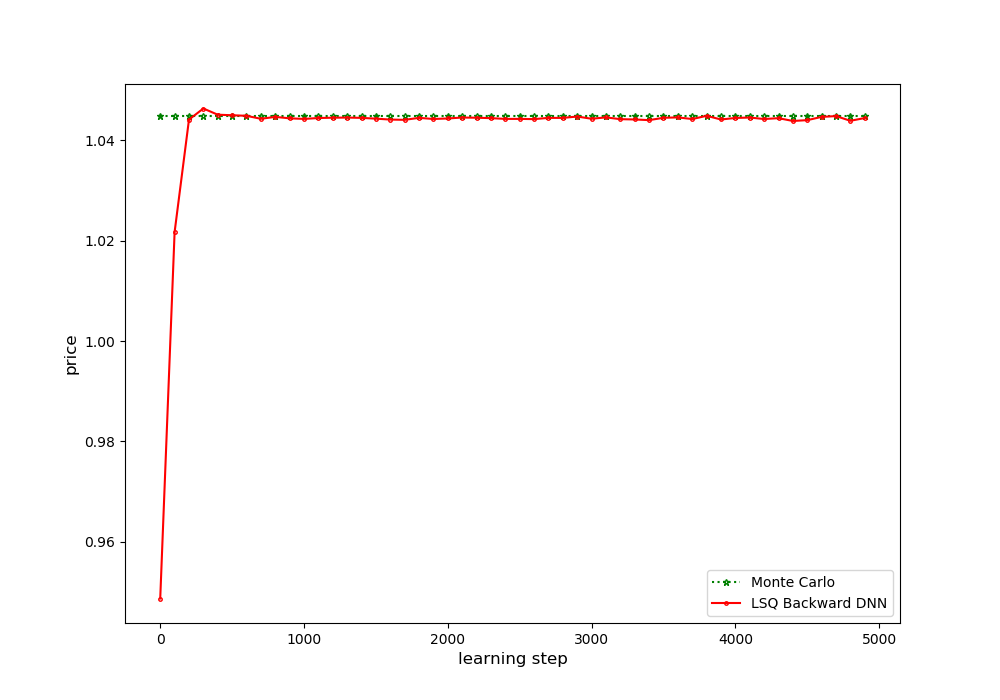}\includegraphics[width=0.5\textwidth]{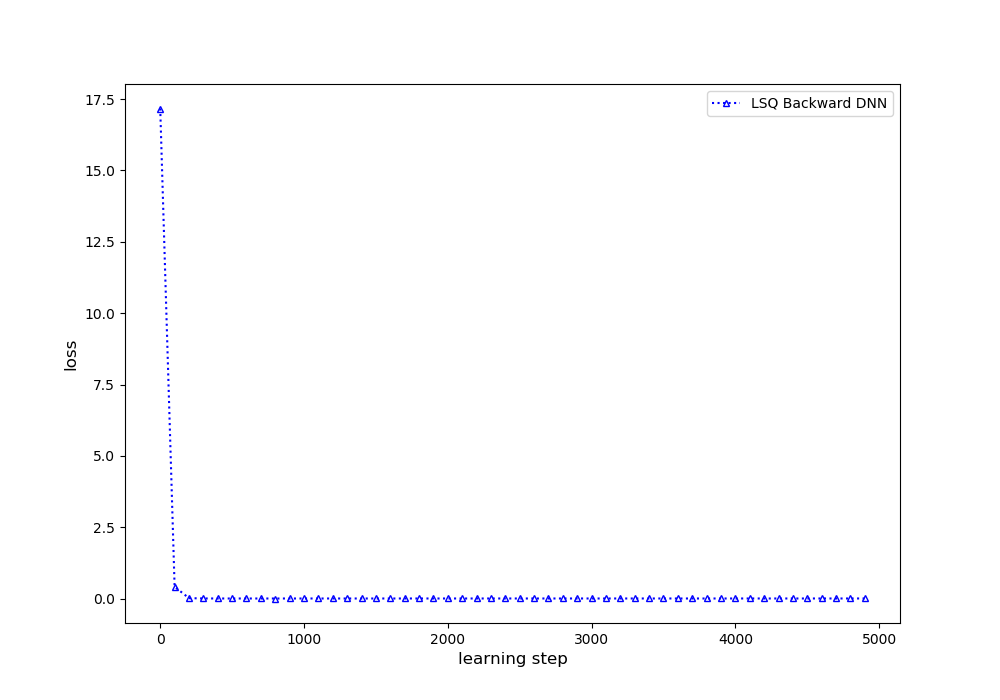}\vspace{-0.5cm}
\caption{CYN, 5 underlying stocks\label{fig:CYN_UL5}}
\end{figure}

In this section, we use 1Y CYNs to test the performance of our least
square backward DNN method for complex payoffs. We test CYNs on a
single underlying stock, 2 underlying stocks (stock 1, 2), 3 underlying
stocks (stock 1, 2, 3) and 5 underlying stocks (stock 1, 2, 3, 4,
5). Some key contract parameters of the tested CYNs are provided in Table
\ref{tab:CYN_para}. We compare the prices from PDE and Monte Carlo with the prices from
the least square backward DNN method. The results are presented in Table \ref{tab:CYN}
and Figure \ref{fig:CYN_UL1} - \ref{fig:CYN_UL5}. It can be
seen that all the tested samples converge fast and the differences
in prices between the backward DNN approach and PDE or MC method is
very small. The results for 10, 20 and 50 dimensions are presented in Section \ref{sec:eff_test_BermCNY}. 
Overall, very accurate prices can be obtained with the least square backward DNN. Similar to Bermudan options, the largest difference
between LSQ MC and the least square backward DNN method occurs at 50 dimensions with a difference of 1.5\% (or 1.5 cents).
The results indicates the validity and accuracy of the least square backward DNN method even when the payoff is complex.

\begin{table}[!th]
\noindent \centering{}\caption{CYN contract parameters\label{tab:CYN_para}}
\begin{tabular}{|c|c|}
\hline 
contingent coupon & $r_{i}=5\%$\tabularnewline
\hline 
coupon barrier & $B_{i}=70\%$\tabularnewline
\hline 
knock-in barrier & $B=50\%$\tabularnewline
\hline 
knock-in put strike & $K=100\%$\tabularnewline
\hline 
call/coupon schedule & quarterly or 0.25, 0.5, 0.75, 1.0\tabularnewline
\hline 
\end{tabular}
\end{table}

\begin{table}[!th]
\noindent \centering{}\caption{Comparison among PDE, Monte Carlo and least square backward DNN method
on CYNs\label{tab:CYN}}
\vspace{-0.3cm}
\begin{center}
\resizebox{.99\textwidth}{!}{
\begin{tabular}{|c|c|c|c|c|c|}
\hline 
\multicolumn{1}{|c|}{{\small{}1Y CYN}} & {\small{}PDE} & {\small{}LSQ MC} & {\small{}LSQ BDNN} & {\small{}rel diff from PDE} & {\small{}rel diff from LSQMC}\tabularnewline
\hline 
\hline 
{{\small{}single stock (stock 1)}} & {\small{}1.0475} & {\small{}1.0474} & {\small{}1.0474} & {\small{}-0.01\%} & {\small{}0.00\%}\tabularnewline
\hline 
{{\small{}2 stocks (stock 1, 2)}} & {\small{}1.0457} & {\small{}1.0458} & {\small{}1.0465} & {\small{}0.08\%} & {\small{}0.07\%}\tabularnewline
\hline 
{{\small{}3 stocks (stock 1, 2, 3)}} & {\small{}1.0438} & {\small{}1.0453} & {\small{}1.0452} & {\small{}0.13\%} & {\small{}-0.02\%}\tabularnewline
\hline 
{{\small{}5 stocks (stock 1, 2, 3, 4, 5)}} & {\small{} } & {\small{}1.0449} & {\small{}1.0448} & {\small{} } & {\small{}0.00\%}\tabularnewline
\hline 
\end{tabular}
}
\end{center}
\end{table}

\subsection{Efficiency test}\label{sec:eff_test}
In this section, we use a 1Y European option, a 1Y Bermudan option and a 1Y CYN to compare
the computation efficiency between the DNN method and classical Monte
Carlo method (least square Monte Carlo for Bermudan options and CYNs). 
We select 5 underlying stocks (stock 1, 2, ..., 5), 10 underlying
stocks (stock 1, 2, ..., 10), 20 underlying stocks (repeat the 10 stocks twice) and 50 underlying stocks 
(repeat the 10 stocks five times) in our tests. The European/Bermudan option contract characteristics are analogous 
to those in Section \ref{sec:Bermudan_Test}. The strike is chosen as $\sum_{d}\omega_{i}X_{0}^{i}$
with equal weight $\omega_{i}=1/d$ so that the option is ATM. The
Bermuda option can be exercised quarterly, or at $t = 0.25, 0.5, 0.75, 1.0$.
The CYN contract characteristics are analogous to those in Section \ref{sec:CYN_test}.
The same parameters are used. For the classical MC (either the regular classical MC or the least square MC) 
we use $M=1,000,000$ sampling paths to estimate the means. 
For the least square backward DNN solver, the MC sampling size is 5,000 and 500 optimization iterations of training 
and validate the trained DNN every 10 iterations. This produces 50 results. We use the mean(standard error) of the 10 results with 
the least loss function value as our derivative price(standard error).

We performed our tests on both a desktop and a server. 
The testing desktop has CPU (Intel(R) Xeon(R) Silver 4108 @1.80GHz) with 8 cores/16 threads and 24GB RAM. 
The sever has 72 cores and 768GB RAM and each core is Intel(R) Xeon(R) E5-2699 v3 @2.30GHz.

It is well known that the parallelization of the least square Monte Carlo,
widely used in high-dimensional American/Bermudan option pricing in
practice, is a challenging task as the regression consumes
most of the computation time for American options and Bermuda options
with many early exercise times. However, since regression at each
exercise date requires the cross-sectional information from all paths,
regression step is not straightforward to parallelize. Realizing this
characteristics, \cite{Choudhury} parallelized the
singular value decomposition in the regression step. Together with
path generation parallelized, an efficient ratio (speed up factor/number
of processors) of 56\% is achieved on a IBM Blue Gene. \cite{Chen2015} 
proposed to apply space decomposition to both the path generation
phase and the regression/valuation phases. In Chen {\itshape et al.}'s work each
sub-sample is an independent least square MC run. The authors found
that the speed-up efficiency can be around 100\% for 8 processes and
around 80\% for 64 processes. Even though there is significant improvement
in parallelization efficiency, pricing bias is observed and the magnitude
of the bias increases with the increase in the number of processes. 
Therefore, instead of using sub-sampling as the parallelization strategy, 
we used the least square regression routine in TensorFlow to implement the regression 
step in the classical least-square MC after the MC samples are generated.  
This choice is in the same spirit as the approach from Choudhury et al., i.e. parallelizing 
the most time consuming component of the process.  Since the classical least-square MC is our benchmark 
to assess the accuracy of our backward DNN method, our approach avoids potential sub-sampling bias 
in the results from classical least-square MC.  In addition, since the computational resources are 
fully managed by TensorFlow for both classical least-square MC and backward DNN, 
we have a fair comparison of efficiency for the two approaches.  

\subsubsection{Efficiency tests on European options}
The testing results for a European option from 5 dimensions to 50 dimensions are presented in Table \ref{tab:Euro_efficiency}.  
It can seen that backward DNN method can produce prices very close (around 2 cents or less) to those from classical MC. 
Though classical MC is faster, it can not produce results for 20 dimensions and above with a desktop due to memory issues.   
The backward DNN method is slower, but results can be produced with all the cases we tested since it does 
not need a large number of samples to produce accurate results. 
Apparently, in general, if one has sufficient computational resource, DNN method is not efficient to price European style options 
as it is 5 to 6 times slower than the classical MC, primarily caused by the high cost in the DNN initialization and optimization.  
However, if one only has a machine with limited memory, the DNN approach is a choice for large scale problems.

\begin{table}[!th]
\noindent \centering{}\caption{Comparison between Monte Carlo and backward DNN method
on European options\label{tab:Euro_efficiency}}
\vspace{-0.3cm}
\begin{center}
\resizebox{.99\textwidth}{!}{
\begin{tabular}{|c|c|c|c|c|c|c|c|c|}
\hline
\multicolumn{1}{|r|}{\multirow{2}{*}{\begin{tabular}[c]{@{}c@{}}1Y ATM\\ European Call\end{tabular}}} & \multicolumn{4}{c|}{MC}                                         & \multicolumn{4}{c|}{BDNN}                                   \\ \cline{2-9} 
\multicolumn{1}{|r|}{}                                                                                & \multicolumn{1}{c|}{price} & std error & desktop(s) & server(s) & \multicolumn{1}{c|}{price} & std error & desktop(s) & server(s) \\ \hline
5 stocks                                                                                              & 8.1033                     & 0.0142    & 78         & 57        & 8.1146                     & 0.0169    & 486        & 330       \\ \hline
10 stocks                                                                                             & 7.2546                     & 0.0127    & 157        & 112       & 7.2318                     & 0.0159    & 882        & 544       \\ \hline
20 stocks                                                                                             & 6.8038                     & 0.0119    & -          & 230       & 6.7856                     & 0.0144    & 1904       & 981       \\ \hline
50 stocks                                                                                             & 6.5121                     & 0.0113    & -          & 576       & 6.4975                     & 0.0137    & 6290       & 2650      \\ \hline
\end{tabular}
}
\end{center}
\end{table}

\subsubsection{Efficiency tests on Bermudan options and CYNs} \label{sec:eff_test_BermCNY}
The testing results from 5 dimensions to 50 dimensions are presented in 
Table \ref{tab:Berm_efficiency} for Bermudan options and in Table \ref{tab:CYN_efficiency} for CYNs. 
First it can be seen that the backward DNN method (here the least square backward DNN) 
can produce prices very close (round 1 cent or less in most cases) to those from classical least square MC, 
an evidence of its validity and accuracy.
It is also very interesting of note that for Bermudan options, the standard deviations of the results from backward DNN 
are very similar to the MC errors of the mean from classical MC approach with our choice of parameters. For CYNs, though 
the standard deviations of the results from backward DNN are one order of higher than the MC errors of the mean from classical MC approach, 
they are still sufficiently small. Similar to the efficient tests for European options,  the classical MC approach, 
though 5 or more times faster,could not produce results for 20 dimensions and above with a desktop computer due to memory issues.

\begin{table}[!th]
\noindent \centering{}\caption{Comparison between Monte Carlo and least square backward DNN method
on Bermundan options\label{tab:Berm_efficiency}}
\vspace{-0.3cm}
\begin{center}
\resizebox{.99\textwidth}{!}{
\begin{tabular}{|c|c|c|c|c|c|c|c|c|}
\hline
\multicolumn{1}{|r|}{\multirow{2}{*}{\begin{tabular}[c]{@{}c@{}}1Y ATM\\ Bermudan Call\end{tabular}}} & \multicolumn{4}{c|}{LSQ MC}                                         & \multicolumn{4}{c|}{LSQ BDNN}                                   \\ \cline{2-9} 
\multicolumn{1}{|r|}{}                                                                                & \multicolumn{1}{c|}{price} & std error & desktop(s) & server(s) & \multicolumn{1}{c|}{price} & std error & desktop(s) & server(s) \\ \hline
5 stocks                                                                                              & 8.2709                     & 0.0124    & 82         & 60        & 8.2795                     & 0.0160    & 509        & 345       \\ \hline
10 stocks                                                                                             & 7.4112                     & 0.0110    & 177        & 116       & 7.4127                     & 0.0153    & 972        & 569       \\ \hline
20 stocks                                                                                             & 6.9760                     & 0.0103    & -          & 237       & 6.9745                     & 0.0141    & 2549       & 976       \\ \hline
50 stocks                                                                                             & 6.7372                     & 0.0100    & -          & 658       & 6.7100                     & 0.0137    & 21552      & 3476      \\ \hline
\end{tabular}
}
\end{center}
\end{table}

\begin{table}[!th]
\noindent \centering{}\caption{Comparison between Monte Carlo and least square backward DNN method
on CYNs\label{tab:CYN_efficiency}}
\vspace{-0.3cm}
\begin{center}
\resizebox{.99\textwidth}{!}{
\begin{tabular}{|c|c|c|c|c|c|c|c|c|}
\hline
\multirow{2}{*}{1Y CYN} & \multicolumn{4}{c|}{LSQ MC}                     & \multicolumn{4}{c|}{LSQ BDNN}               \\ \cline{2-9} 
                        & price  & std error & desktop(s) & server(s) & price  & std error & desktop(s) & server(s) \\ \hline
5 stocks                & 1.0449 & 0.0000    & 84         & 60        & 1.0448 & 0.0008    & 506        & 342       \\ \hline
10 stocks               & 1.0402 & 0.0001    & 179        & 122       & 1.0390 & 0.0011    & 947        & 563       \\ \hline
20 stocks               & 1.0257 & 0.0001    & -          & 235       & 1.0236 & 0.0016    & 2860       & 1012      \\ \hline
50 stocks               & 0.9778 & 0.0002    & -          & 659       & 0.9633 & 0.0023    & 20947      & 3274      \\ \hline
\end{tabular}
}
\end{center}
\end{table}

\section{Conclusion\label{sec:Conclusion}}

In this work, we have developed a deep learning-based least square
forward-backward stochastic differential equation solver, which can
be used in high-dimensional derivatives pricing. Our deep learning
implementation follows a similar approach to the ones explored by
\cite{E2017,Han8505} and \cite{Haojie}.
However, the forward DNN method is more suitable for European style
derivative pricing and the backward DNN method from Wang {\itshape et al.} may
not adequately account for  early exercise features. In our approach,
we embed the least square regression technique similar to that in
the least square Monte Carlo method (\cite{LSQ_MC}) to the backward
DNN algorithm. Numerical testing results on Bermudan options and callable
yield notes indicate that our least square backward DNN method can
produce very accurate results based on comparisons with the finite difference based PDE approach and the classical Monte Carlo simulation. 
This method can be used in various derivative pricing applications such as
Barrier option, American option, convertible bonds, etc. In conclusion,
our least square backward DNN algorithm can serve as a generic numerical
solver for pricing derivatives, and it is most suitable for high-dimensional
derivatives with early exercises features. Though it is slower than the classical Monte Carlo, the backward 
DNN approach is very memory efficient and can handle large scale problems with less computational resources.

\section*{Acknowledgements }

The authors would like to
thank Dr. Agus Sudjianto for introducing them to the field
of using machine learning on derivative pricing and thank Prof. Liuren
Wu for reviewing the manuscript. The authors also gratefully acknowledge Dr. Bernhard Hientzsh 
for highly valuable discussions.

\end{document}